\documentclass[12pt]{article}
\usepackage{amsthm,url}
\usepackage{amsmath}
\usepackage{amssymb}
\usepackage{color}
\usepackage[square,numbers,comma,sort&compress]{natbib}
\usepackage{enumerate}
\usepackage{colortbl}
\usepackage{booktabs}
\usepackage{ifpdf}
\usepackage{graphics}
\usepackage{epsfig}
\usepackage[TABBOTCAP,normalsize]{subfigure}
\usepackage{booktabs}
\usepackage{mathrsfs}  
\usepackage{rotating} % For sideways text in tables

\usepackage{times}

\usepackage{sectsty}

\sectionfont{\large}
\subsectionfont{\large}

\newtheorem{definition}{Definition}

\newtheorem{proposition}{Proposition}
\newtheorem{algorithm}{Algorithm}

\newcommand\R{\mathbb R}  
\newcommand\Side[1]{\begin{sideways}{\small #1}\end{sideways}}

\title{Statistical Phylogenetic Tree Analysis Using Differences of Means}
\begin{document}

%\maketitle
\begin{center}
{\large
Statistical Phylogenetic Tree Analysis Using Differences of Means
}
\end{center}
\vskip 5pt
\begin{center}
    Elissaveta Arnaoudova\footnote{EA and DH contributed equally to this work}\footnote{Department of Computer Science, University of Kentucky}, David Haws\footnotemark[1]\footnote{Department of Statistics, University of Kentucky}, Peter Huggins\footnote{Lane Center for Computational Biology, Carnegie Mellon},\\ Jerzy W. Jaromczyk\footnotemark[2], Neil Moore\footnotemark[2], Chris Schardl\footnote{Department of Plant Pathology, University of Kentucky}, Ruriko Yoshida\footnotemark[3]\footnote{Ruriko Yoshida,
University Of Kentucky, Department of Statistics, 817 PATTERSON OFFICE TOWER, LEXINGTON KY  40506-0027, ruriko.yoshida@uky.edu}
\end{center}
\begin{abstract}
We propose a statistical method to test whether two phylogenetic trees with given alignments are significantly incongruent.  Our method compares the two distributions of phylogenetic trees given by the input alignments, instead of comparing point estimations of trees. This statistical approach can be applied to gene tree analysis for example, detecting unusual events in genome evolution such as horizontal gene transfer and reshuffling. Our method uses difference of means to compare two distributions of trees, after embedding trees in a vector space. Bootstrapping alignment columns can then be applied to obtain p-values.  To compute distances between means, we employ a ``kernel trick'' which speeds up distance calculations when trees are embedded in a high-dimensional feature space, e.g. splits or quartets feature space.  In this pilot study, first we test our statistical method's ability to distinguish between sets of gene trees generated under coalescence models with species trees of varying dissimilarity. We follow our simulation results with applications to various data sets of gophers and lice, grasses and their endophytes, and different fungal genes from the same genome. A companion toolkit, {\tt Phylotree}, is provided to facilitate computational experiments.
%In addition to theoretical analysis of the method, a companion toolkit, {\tt Phylotree}, is provided to facilitate computational experiments. Given input alignments, it samples the corresponding phylogenetic trees using, e.g., {\tt BEAST} and {\tt MrBayes}, and then computes distances between means of posterior tree distributions, for both original and bootstrapped sequences.
\end{abstract}

\section{Introduction}

Estimating differences between phylogenetic trees is one of the fundamental questions in computational biology.  Conflicting phylogenies arise when, for example, different phylogenetic reconstruction methods are applied to the same data set, or even with one reconstruction method applied to multiple different genes. Gene phylogenies may be codivergent by virtue of congruence (identical trees) or insignificant incongruence. Otherwise, they may be significantly incongruent \cite{Maddison1997}. All of these outcomes are fundamentally interesting. Congruence of gene trees (or subtrees) is often considered the most desirable outcome of phylogenetic analysis, because such a result indicates that all sequences in the clade are orthologs (homologs derived from the same ancestral sequence without a history of gene duplication or lateral transfer), and that discrete monophyletic clades can be unambiguously identified, perhaps supporting novel or previously described taxa. In contrast, gene trees that are incongruent are often considered problematic because the precise resolution of speciation events seems to be obscured. Thus, it would also be very useful to identify significant incongruencies in gene trees because these represent noncanonical evolutionary processes.  (e.g., \cite{Maddison2006, Edwards2007, Liu2008, Liu2007}).  
In this paper we propose a statistical hypothesis test which tells whether two phylogenetic trees are significantly incongruent to each other by comparing two distributions for phylogenetic trees, instead of comparing two point estimations.  More specifically we will compare two distributions of trees using {\em difference of means}.  In this paper we estimate these distribution by Bayesian sampling from the posterior distribution. Our statistical hypotheses are:
\begin{quote}
$H_0$: Phylogenetic trees $T_1$ and $T_2$ are congruent. \\  
$H_1$: Phylogenetic trees $T_1$ and $T_2$ are incongruent.
\end{quote}
Usually a statistical test on the above hypotheses considers point estimates of the trees obtained by a tree reconstruction method, such as maximum likelihood estimates \citep{Felsenstein1981,Galtier2005} or the neighbor-joining method \citep{Saitou1987}.  See \cite{chris} and references within for an overview.  Variation of reasonable tree estimates can be assessed, for example, by using the bootstrap or jackknife method.

There are several techniques to test if gene trees are codiverged.  
For example, the Bayesian estimation methods (e.g., \citep{Liu2007, Edwards2007, Ane2007}), the Templeton test implemented in {\tt paup*} \citep{Swofford1998}
(e.g., 
\citep{tmp}), the partition-homogeneity test (PHT) also implemented with 
{\tt paup*} (e.g., \citep{PHT}),  Kishino-Hasegawa test (e.g., 
\citep{HK}), and the likelihood ratio test (LRT; e.g., \citep{geneLRT}) are statistical methods to see 
if there is a ``significant'' level of incongruence between the trees (these methods are also called partition likelihood support (PLS) \citep{PLS}). 
However, there is a limitation in many methods for comparing two phylogenetic trees:  It is implicitly assumed that the two given trees are actually correctly estimated phylogenies.  In reality, trees are estimated from observed data (e.g. fossil record, sequence data), and tree uncertainty is the rule instead of the exception.  Thus, to estimate trees, we propose using posterior means instead of maximum likelihood, and we apply the bootstrap method to assess variation in the posterior means.  %Instead of using point estimates we use distributions of trees so that we consider the uncertainty of each tree.  
Our method could also be applied with tree estimators like maximum likelihood, instead of the posterior mean.

This paper is organized as follows:  In Section 2, we state our method.  In Section 3, we show simulation studies with data generated by the software {\tt Mesquite} [Maddison Knowles 2006]. In Section 4, we apply our method to well-known gopher-louse data sets from \citep{Hafner} and grass-endophyte data sets from \citep{chris}. We end with a discussion.

\section{Materials and Methods}\label{Methods}
\subsection{Preliminaries}

Let $\mathcal{T}_n$ be the space of trees on $n$ taxa.  
When analyzing and comparing phylogenies, often {\em tree features} are used.  The notion of tree features can be expressed formally as a vector space embedding:

\begin{definition}
Given a vector space embedding $v:\mathcal{T}_n \to {\mathbb R}^m$ for some $m$, the vector $v(T)$ is the {\em feature vector} of $T$. 
\end{definition}

The difference between trees $T_1,T_2$ can be quantified as the distance $||v(T_1) - v(T_2)||$, where $|| \cdot ||$ is any norm.  In this paper we will focus on $L_2$ norms.

A notable example of our framework is the {\em dissimilarity map distance}.  

\begin{definition}
For $T \in {\mathcal T}_n$, let $v(T) = (d_{1,2}^T, d_{1,3}^T, \ldots, d_{n-1,n}^T) \in {\mathbb R}^{n \choose 2}$ be the vector of pairwise distances $d_{i,j}^T$ between leaves $i$ and $j$ in $T$.  The {\em dissimilarity map distance} is
$$d(T_1,T_2) = ||v(T_1) -  v({T_2})|| = \sqrt{(d_{1,2}^{T_1} - d_{1,2}^{T_2})^2 + \ldots + (d_{n-1,n}^{T_1} - d_{n-1,n}^{T_2})^2}$$
where $|| \cdot ||$ represents $L_2$ norm (Euclidean length).  
%$||v(T_1) -  v({T_2})||$ is the Euclidean length of the vector $v(T_1) -  v({T_2})$, and it is calculated by the root-square of sum of squares in each coordinate of the vector based on the Pythagorean Theorem.
\end{definition}

In our computational experiments, we will use the dissimilarity map distance.
Dissimilarity map distance was studied in \citep{Buneman1971}.  One can also consider a variation where all edge lengths are set to 1.  The arising dissimilarity distance is called the {\em path difference} and only depends on tree topologies.

\subsection{Testing for congruence of two trees}

In our framework, given are $D_1, \,D_2$, each a collection of $n$ aligned homologous sequences.  We assume $D_1, \, D_2$ were generated by models of sequence evolution on unknown trees $T_1, \,T_2$.  After embedding trees into a vector space, our statistical hypotheses are:
\begin{quote}
$H_0$: $||v(T_1) - v(T_2)|| = 0$, \\
$H_1$: $||v(T_1) - v(T_2)|| > 0$.
\end{quote}

For convenience, we describe our approach as comparing two gene trees $T_1, \, T_2$ from the same set of species.  One can also compare a phylogeny for host species and a phylogeny for corresponding parasites, as we do in section \ref{realdata}. 

Random fluctuations in sequence evolution can cause reconstructed gene trees for $D_1$ and $D_2$ to look at least slightly different, even if the true underlying trees are equal.  Thus we need a way to tell if the difference between two estimated trees is "significant.''

One classical approach to assess variability in reconstructed trees is the bootstrap \cite{Felsenstein1981}.  The bootstrap generates new hypothetical sequence alignments, by sampling (with replacement) columns of aligned sequence.  Then trees can be re-estimated for each hypothetical alignment.  One common application of the bootstrap is to measure support for each clade; clades that appear in most bootstrap replicate trees are regarded as likely clades in the true tree.

Here we propose a bootstrap procedure to assess significance of the distance between two trees.  Our method is based on the triangle inequality.  Namely, if $v(\hat{T_1}), \, v(\hat{T_2})$ are estimators for $v(T_1), \, v(T_2)$, then the triangle inequality says
\[
    ||v(T_1) - v(T_2)|| \geq ||v(\hat{T_1}) - v(\hat{T_2})|| - ||v(T_1) - v(\hat{T_1})|| - ||v(T_2) - v(\hat{T_2})||,
\]
which gives a lower bound on the distance between the true trees $T_1, \, T_2$.   See Figure 1 for an illustration.  We cannot compute the right-hand side of the inequality directly, because $T_1, \, T_2$ are unknown.  Instead, we use the bootstrap to estimate the distributions of the terms $||v(T_1) - v(\hat{T_1})||$ and $||v(T_2) - v(\hat{T_2})||$.   An outline of our bootstrap procedure is in the Supplement.

%\begin{figure}[!htp]
%\begin{center}
%\includegraphics[height=3.5cm]{DiffMeans.jpg}
%\end{center}
%\caption{A diagram showing two cases of the differences of means method. In the first diagram the distance between $v(T_1)$ and $v(T_2)$ ($d_1$) is greater than the distance between $v(T_1)$ and $v(\widehat{T_1})$ ($d_2$) plus the distance between $v(T_2)$ and $v(\widehat{T_2})$ ($d_3$). i.e., $d_1 \geq d_2 + d_3$. In the second figure we see $d_1 \leq d_2 + d_3$. }
%\end{figure}

\begin{figure}[!htp]
\begin{center}
\subfigure[]{
\includegraphics[height=3.5cm]{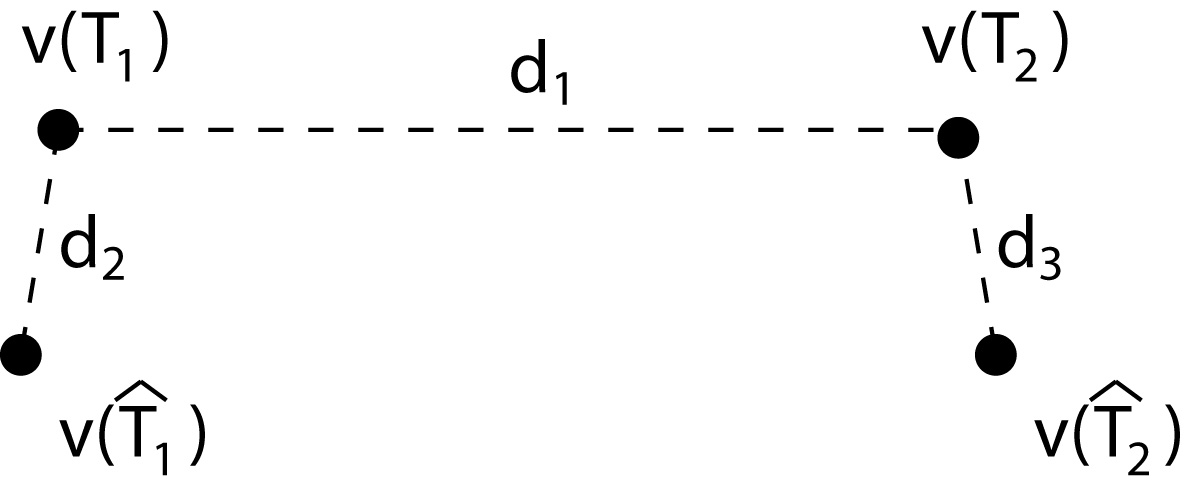}
\label{subfig:diffmeans_far}
}
\subfigure[]{
\includegraphics[height=3.5cm]{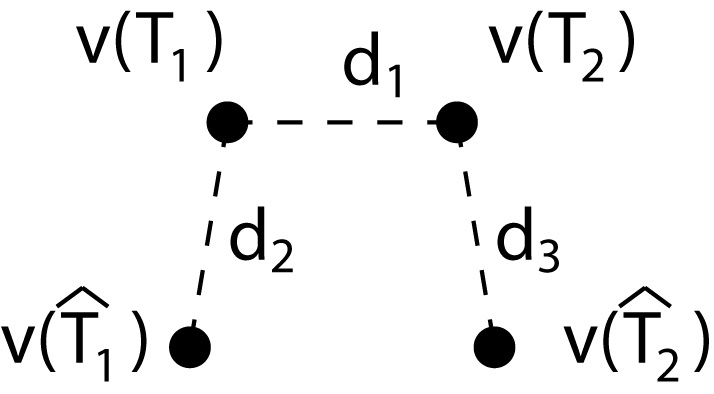}
\label{subfig:diffmeans_short}
}
\end{center}
\caption{A diagram showing two cases of the differences of means method. In Subfigure \ref{subfig:diffmeans_far} the distance $d_1$ between $v(T_1)$ and $v(T_2)$  is greater than the distance $d_2$ between $v(T_1)$ and $v(\widehat{T_1})$ plus the distance $d_3$ between $v(T_2)$ and $v(\widehat{T_2})$, i.e., $d_1 \geq d_2 + d_3$. In Subfigure \ref{subfig:diffmeans_short} we see $d_1 \leq d_2 + d_3$. }
\end{figure}

\subsection{Difference of means}

The bootstrap procedure we have proposed can be applied with any tree estimator, such as neighbor joining or maximum likelihood.  Since we are presuming tree uncertainty is high, and Bayes estimator trees are more accurate than neighbor joining or ML \cite{BEpaper}, we prefer a Bayes estimator approach.  

Given an alignment $D$, generated by sequence evolution on an unknown tree $T$, Bayesian MCMC sampling methods will approximately sample from the posterior distribution $P(T \, | \, D) \sim P(D \, | \, T ) P(T)$ \cite{Yang1997}.    For two posterior distributions $P(T_1 \, | \, D_1)$ and $P(T_2 \, | \, D_2)$, let $t_1, \ldots, t_{N_1}$ be a sample from $P(T_1 \, | \, D_1)$, and similarly for $t_1', \ldots, t_{N_2}'$ a sample from $P(T_2 \, | \, D_2)$.  Then we can use $\frac{1}{N_1} \sum_{i=1}^{N_1} v(t_i)$ as an estimator for $v(T_1)$, and similarly for $v(T_2)$.
The {\em difference of means} is   

\begin{equation}\label{delta}
\hat{\Delta} = \frac{1}{N_1} \sum_{i=1}^{N_1} v(t_i) - \frac{1}{N_2} \sum_{i=1}^{N_2} v(t_i')
\end{equation}

and $|| \hat{\Delta} ||$ is an estimator for $||v(T_1) - v(T_2)||$.

\subsubsection{The kernel trick for estimating $|| \hat{\Delta} ||$}

Some feature space embeddings produce very high-dimensional feature vectors $v(T_1), v(T_2)$, yet the distance $||v(T_1) - v(T_2)||$ can be computed quickly without explicitly writing down the feature vectors for $T_1$ and $T_2$. 
Notable examples include Robinson-Foulds distance and quartet distance.  In such cases, it would be desirable if the difference of means $|| \hat{\Delta}||$ could be estimated, by sampling trees and computing the distances between samples (without writing down any feature vectors).  This is indeed possible, using a {\em kernel trick}:  

\begin{proposition}\label{prop1}
Let $x_1, x_2, y_1, y_2 \in {\mathbb R}^m$ be four pairwise independent random variables, where $x_1$ and
$x_2$ are drawn according to distribution $P$, and $y_1$, $y_2$ are drawn
according to distribution $Q$ such that ${\mathbb E}(x_1) = {\mathbb E}(x_2) = \mu_x \in {\mathbb R}^m$ and ${\mathbb E}(y_1) = {\mathbb E}(y_2) = \mu_y \in {\mathbb R}^m$.  Then

\begin{equation}\label{diffmeans}
%\begin{array}{lrl} 
|| \mu_x  - \mu_y || ^2  =   {\mathbb E}( || x_1 - y_1 ||^2 ) -\frac{1}{2}\left[{\mathbb E}( || x_1 - x_2 ||^2 )\right] -\frac{1}{2}\left[{\mathbb E}( || y_1 - y_2 ||^2 )\right]. 
%\end{array}
\end{equation}
\end{proposition}

A proof of Proposition \ref{prop1} is provided in the supplement. Using the proposition, and a subroutine which can compute $||v(t) - v(t')||$ for two given tree samples $t, t'$, the length $|| \hat{\Delta} || = ||{\mathbb E}v(T_1) - {\mathbb E}v(T_2)||$ can be estimated from the samples $\{t_i\}, \{t_i'\}$.  

\section{Results}

\subsection{Simulations}

%The difference of means method attempts to test if two distributions are different.  
In this section we estimate posterior distributions of phylogenetic trees via MCMC-based software {\tt MrBayes} \cite{Mrbayes} and apply the difference of means method to test if two phylogenetic trees are incongruent.  Note other softwares such as {\tt BEAST} \cite{beast} could also be used.  Simulated data sets were generating using the software {\tt Mesquite} \cite{Maddison2006} with parameters chosen similar to  \cite{Maddison2006}, to emulate real data and test the effectiveness of our method.  {\tt Mesquite} takes two parameters; the species depth in terms of number of generations and the population size in terms of number of individuals.  Three simulation sets were generated, determined by the species depths of $100000$, $600000$, and $1000000$. The effective population size was fixed to $100000$ for all data sets. For each simulation set, two species trees, species tree one and two, with eight species were generated using the pure birth yule process in {\tt Mesquite}. Sequence alignments were generated by {\tt Mesquite} under HKY85 model with transition-transversion ratio of 3.0, a discrete gamma distribution with four categories and shape parameters 0.8. In all our simulations, we set the stationary probability distribution $\pi = (0.3, 0.2, 0.2, 0.3)$ for A, C, G, T respectively, the 3:2 AT:GC ratio was maintained through all trees, and our sequences were generated with $1000$ base pairs. The coalescence gene trees generated had branch lengths in terms of the coalescence model and therefore a scaling factor of $10^{-8}$ was used to yield sequences with sequence divergence similar to real data.  Table \ref{seq_div} shows sequence divergences. The sequence divergence was calculated in two ways: (1) the average percent pairwise difference between all sequences \cite{Maddison2006}, and (2) the minimum of the pairwise percent differences among sequences \cite{phyml}.

\begin{table}[!htp]
\subtable[Pairwise Minimum]{
\label{fig:seqdivmin}
%\begin{center}
%Min
\begin{tabular}{l|ccccc}
\Side{\begin{minipage}{1.7cm}Species\\ Depth\end{minipage}} & Min & Q1 & Median & Q3 & Max \\
\midrule
1000K &     0.000  & 0.002 & 0.005 & 0.008 & 0.017\\
600K  &      0.000 & 0.003 & 0.006 & 0.01  & 0.022\\
100K  &      0.000 & 0.001 & 0.001 & 0.002 & 0.006\\
\end{tabular}
%\end{center}
}
\subtable[Pairwise Average]{
\label{fig:seqdivavg}
%\begin{center}
%Avg
\begin{tabular}{l|ccccc}
\Side{\begin{minipage}{1.7cm}Species\\ Depth\end{minipage}} & Min & Q1 & Median & Q3 & Max \\
\midrule
1000K SD &    0.032  &  0.04   & 0.043  &  0.045  &  0.054\\
600K SD  &     0.025 &  0.03   & 0.032  &  0.035  &   0.046\\
100K SD  &     0.004 &  0.007  & 0.008  &  0.012  &  0.016\\
\end{tabular}
%\end{center}
}
\caption{Q1 means the first quantile and Q3 means the third quantile.  By ``min'' we mean the smallest number and ``max'' means the largest number among a sample.  Sequence divergences were calculated in two ways: 1) the pairwise minimum percentage of sequence divergence and 2) the average pairwise percentage of sequence divergence.}\label{seq_div}  
\end{table}

\begin{figure}[!htp]
\begin{center}
\includegraphics[width=18cm]{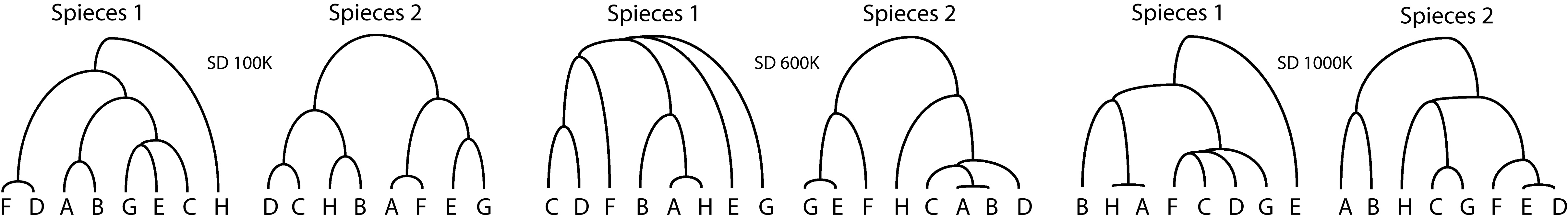}
\end{center}
\caption{The three pairs of species trees used in our simulations.    The dissimilarity maps normalized by $\sqrt{n \choose 2}$ between the two species trees used to generate gene trees for our simulations are $0.4333$ for $1,000,000$ species depth, $0.2672$ for $600,000$ species depth and $0.046$ for $100,000$ species depth.  }\label{species} 
\end{figure}

In order to estimate posterior distributions we used the MCMC-based software {\tt MrBayes} with the following parameters: 
(1) for the model: HKY85 + Gamma, shape parameter: $0.8$, transition-transversion ratio: $3.0$; and (2) for MCMC runs: number of runs: 1, number of chains: $2$, chain length: $100,000$, sample frequency: $1,000$, burn-in: $25$\%.
For bootstrap sampling we sampled $100$ bootstrap samples with sample size of $1,000$ columns since the simulated sequences are generated with $1,000$ base pairs.

\begin{figure}[!htp]
\subfigure[Box plots of p-values for simulated data.  We generated ten gene trees for each species depth (i.e. $30$ different gene trees in total) for each species tree.  Sequences for each gene tree were generated using the HKY model.  Species depths of $1,000,000$, $600,000$ and $100,000$ were used, with fixed population size of $100,000$.  White box plots are for testing Type I error by comparing identical gene trees generated by {\tt Mesquite} (we used the species tree on the right in Figure \ref{species}), light grey box plots represent the p-values with two different gene trees generated from the same species tree  under the coalescence model  (we used the species tree on right in Figure \ref{species}), and dark grey box plots summarize the p-values with gene trees generated from two different species trees.   Some boxes of p-values are identically zero.  Box plots were  computed in {\tt R} \cite{R} using {\tt boxplot}]{
\label{boxplots}
%\begin{center}
$\phantom{AAAAAAAAAAAA}$
\includegraphics[width=9cm]{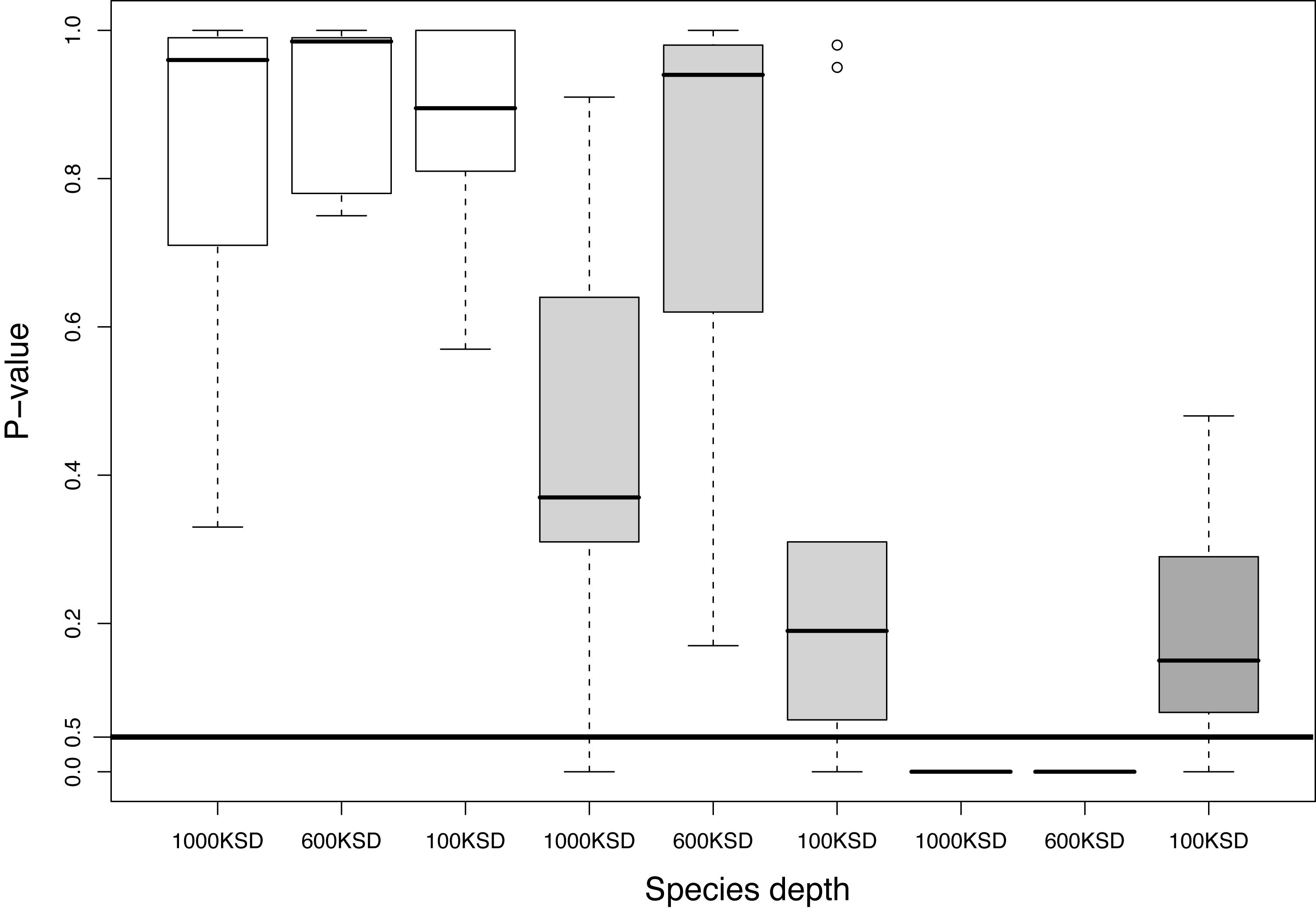}
$\phantom{AAAAAAAAAAAA}$
%\end{center}
%\label{boxplots} 
}\\
%\caption{We tested ten gene trees for each species depth (i.e. 30 different gene trees in total) generated under the same species tree.  We used two sets of sequences generated under the HKY model with the same tree for each test.  We have the three species depths of 1000,000, 600,000 and 100,000, with fixed population size of 100,000.  White box plots are for testing Type I error by comparing the identical gene trees generated by {\tt Mesquite} (we used the species tree on right with each species depth in Figure \ref{species}), light grey box plots represents the p-values with gene trees generated from the same species tree  under the coalescence model  (we used the species tree on right with each species depth in Figure \ref{species}), and dark grey box plots summarize the p-values with  gene trees generated from different species trees  under the coalescence model.  For testing Type I error, the sample size is 10 for each species depth.  The sample size is 9 for other tests for each species depth.  Thus there are 94 pairs in total.}
%\end{figure}
%\begin{figure}[!htp]
\subfigure[These plots represent correlation between distances among true trees and p-values, and correlation between difference of means and p-values.  These are computed with the same data sets in Figure \ref{boxplots}.  The sample size here is 94 (i.e., there are 94 pairs of sequence data sets we tested in total).  For more details see supplement.  We fitted the data in {\tt R} \cite{R} using {\tt loess} to perform local regression.  The dotted lines are for 95\% confidence intervals of the fitted lines.  The vertical solid line is the line $x = 0.05$ which represents the $\alpha$-level, p value $= 0.05$.]{
\label{fitting}
%\begin{center}
$\phantom{AAAAAAAAAAAA}$
\includegraphics[width=9cm]{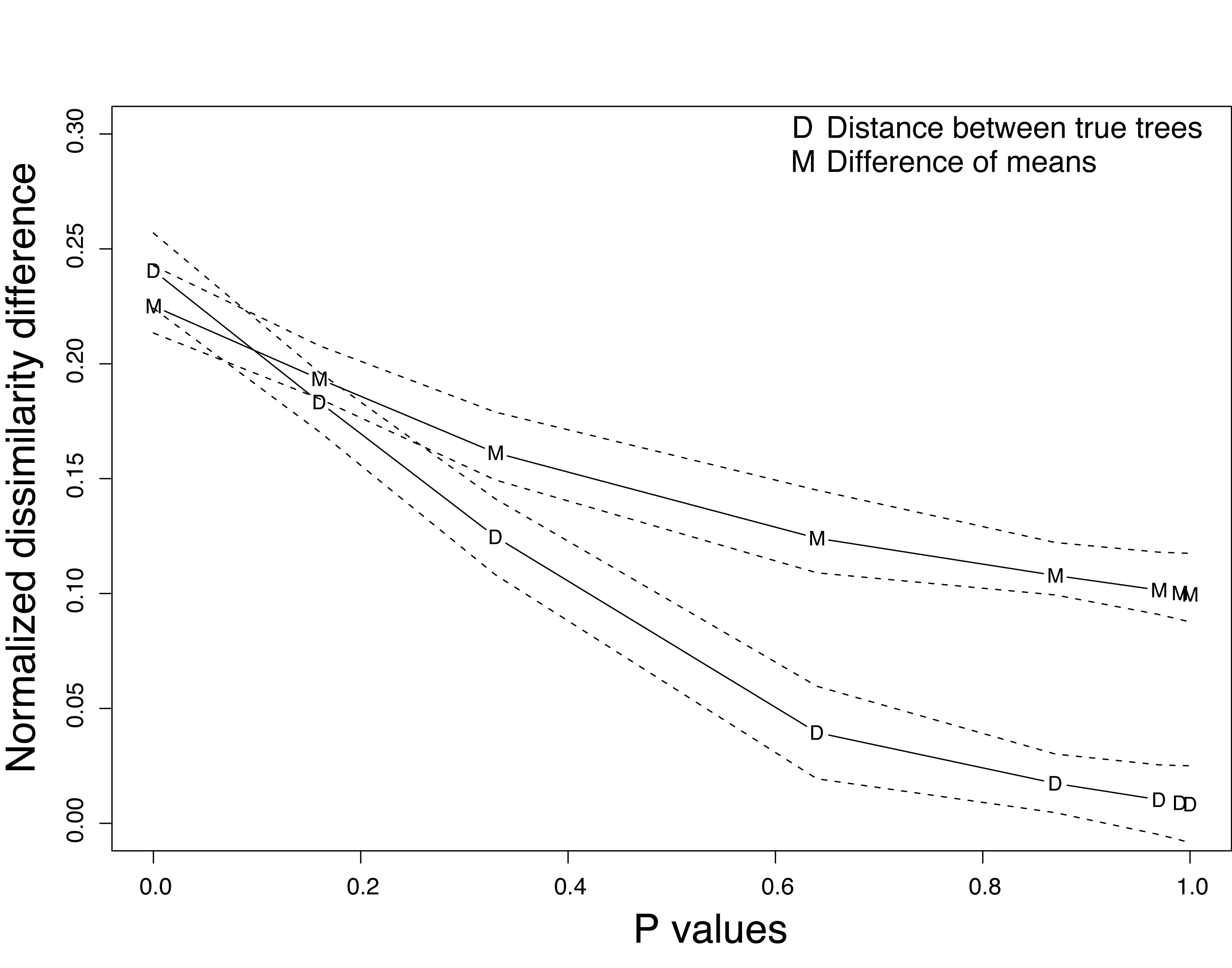}
$\phantom{AAAAAAAAAAAA}$
%\end{center}
%\label{fitting} 
%\caption{this plots represent correlation between distances among true trees and p-values, and correlation between difference of means with the posteriors given the original sequence data sets and p-values.  these are computed with the same data sets in table \ref{boxplots}.   the sample size here is 94 (i.e., there are 94 pairs of sequence data sets we tested in total.  for more details see supplement).  we fitted the data via loess which is a local non-linear regression technique.  the dotted lines are for 95\% confidence intervals of the fitted lines.  the vertical solid line is the line $x = 0.05$ which represents the $\alpha$-level, p value $= 0.05$.}
}
\caption{}
\end{figure}

We generated simulated data sets in three different ways; (1) two separate sequence data sets generated from the same gene tree, (2) sequence data sets generated from two different gene trees under the same species tree, (3) sequence data sets generated by two sequence data sets generated from two different gene trees whose species trees are also different.   We tested ten gene trees for each species depth (i.e. 30 different gene trees in total) generated under the same species tree.  One can find the species trees we used in Figure \ref{species}.  We used two sets of sequences generated under the HKY model with the same tree for each test.  We have the three species depths of 1000,000, 600,000 and 100,000, with fixed population size of 100,000.  Notice that we do not observe any Type I errors with our testing method, however, in within-species comparisons at species depth of 1,000,000 the p-values were high in general (Figure \ref{boxplots}).   Also notice that with pairs of gene trees where each pair of gene trees are generated from different species trees under the coalescence model, the p-values were 0.000 for all pairs of genes from 1,000,000 and 600,000 species depth. However, in the case of species depth 100,000 we see that only one pair (Species1\_Genetree0 / Species2\_Genetree7) has a p-value less than 0.05 (see Table \ref{typeII} in the supplement). We appear to have Type II errors, probably because species trees with 100,000 species depth are very close to each other.

p-values and distance between true trees appear strongly correlated.  We fitted correlations between p-values and distance between true trees as well as correlation between p-values and the difference of means for the posterior distributions given the original sequence data sets, using a function called {\it loess} (Figure \ref{fitting}).   The fitted lines show negative correlation between the  p-values and the distance between true trees and also negative correlation between the p-values and the difference of means.  Note that the fitted lines for distances between true trees and for differences of means in Figure \ref{fitting} any p-values below the $\alpha$-level ($0.05$ in our case) are within their confidence intervals.  Actually they are within their confidence intervals up to the p-value equals to $0.3$.  This means the differences of means with posterior distributions given the original sequence data sets are good measurements for distance between true trees for our statistical tests.  This is particularly important since we usually do not know the true trees with biological data sets. For complete results of our simulations see Table \ref{typeI} and Table \ref{typeII} in the supplement.

\subsection{Experiments with real data sets} \label{realdata}

We tested our method with a well-known gopher-louse data set \cite{Hafner}, see
Table \ref{GL}.  This data set  contains $17$ taxa of lice and $15$ taxa of gophers.   In order to satisfy the requirement for an equal number of leaves for tree comparison we constructed $4$ individual data sets reflecting all possible pairings of the two gopher species involved in the possible host jumps with their apparent parasitic louse species: (dataset 1) Thomomys talpoides-Thomomydoecus barbarae, Thomomys bottae-Thomomydoecus minor; (dataset 2) Thomomys talpoides-Geomydoecus thomomyus, Thomomys bottae-Thomomydoecus minor; (dataset 3) Thomomys talpoides-Thomomydoecus barbarae, Thomomys bottae-Geomydoecus actuosi; (dataset 4) Thomomys talpoides-Geomydoecus thomomyus, Thomomys bottae-Geomydoecus actuosi.

The posterior distributions were estimated using MrBayes with the following parameters: (1) for the model: GTR + Gamma + Invariant sites;
(2) for {MCMC:} number of runs: $1$, number of chains: $2$, chain length: $100,000$, sample frequency: $1,000$, burn-in: $25$\%; and (3) for {bootstrap sampling:} $100$ bootstrap samples with sample size of $379$ columns  which is the length of sequence alignments in the data sets.  

%\begin{table}[!htp]
%\begin{center}
%\begin{tabular}{l|c}
%Data set& p-value  \\
%\midrule
%Gopher-louse (dataset 1) & 0.640\\
%Gopher-louse (dataset 2) & 0.400\\
%Gopher-louse (dataset 3) & 0.840\\
%Gopher-louse (dataset 4) & 0.590\\
%\end{tabular}
%\end{center}
%\label{GL}  
%\caption{We tested our methods with a well-known gopher-louse data set in \cite{Hafner} and this tables shows the p-value for our hypotheses.  All of them are pretty high so that we do not reject the null hypothesis.}
%\end{table}

\begin{table}[!htp]
\begin{center}
\subtable[p-values for subsets  of the well-known gopher-louse data set in \cite{Hafner}.  All p-values are high, so no significant incongruence is found.]{
\label{GL}  
\begin{tabular}{l|c}
Data set& p-value  \\
\midrule
Gopher-louse (dataset 1) & 0.640\\
Gopher-louse (dataset 2) & 0.400\\
Gopher-louse (dataset 3) & 0.840\\
Gopher-louse (dataset 4) & 0.590\\
\end{tabular}
$\phantom{AAAA}$
%\caption{We tested our methods with a well-known gopher-louse data set in \cite{Hafner} and this tables shows the p-value for our hypotheses.  All of them are pretty high so that we do not reject the null hypothesis.}
}
\subtable[p-values for grass-endophyte data sets from \cite{chris}.  After removing cases of apparent host jumps, the data comprises 20 taxa of grasses and 20 taxa of endophytes.   The first two rows compare grass phylogeny to gene trees for {\em tefA} and {\em tubB} in endophytes; the last row uses the concatenation of {\em tefA} and {\em tubB}.]{
\label{PE}  
$\phantom{AAAA}$
\begin{tabular}{l|c}
Data set& p-value  \\
\midrule
Grass-endophyte {\em tefA} & 0.040\\
Grass-endophyte {\em tubB} & 0.080\\
Grass-endophyte {\em tubB} plus {\em tefA} & 0.000\\
\end{tabular}
%\caption{This table shows the p-values for the hypotheses with our statistical Method with the grass-endophyte data sets from \cite{chris}.  After removing cases of apparent host jumps, the data sets contain sequences from 20 taxa of grasses and 20 taxa of endophytes.   The first row represents for the endophyte data sets from {\em tefA} and the second row is from {\em tubB}.  The last row represents for the data sets with a concatenated sequences from {\em tubB} and {\em tefA}.}
}
\caption{}
\end{center}
\end{table}

We also tested our Method with the data sets from \cite{chris}.  After removing cases of apparent host jumps, the data sets contain sequences from 20 taxa of grasses and 20 taxa of endophytes.  Sequences were aligned with the aid of {\tt PILEUP} implemented in {\tt SEQWeb} Version 1.1 with {\tt Wisconsin Package} Version 10 (Genetics Computer Group, Madison, Wisconsin). PILEUP parameters were adjusted empirically; a gap penalty of two and a gap extension penalty of zero resulted in reasonable alignment of intron-exon junctions and intron regions of endophyte sequences, and of intergenic spacer and intron regions of cpDNA sequences. Alignments were scrutinized and adjusted by eye, using tRNA or protein coding regions as anchor points. For phylogenetic analysis of the symbionts, sequences from {\em tubB} (encoding $\beta$-tubulin) and {\em tefA} (encoding translation elongation factor 1-$\alpha$) were concatenated to create a single, contiguous sequence of approximately 1400 bp for each endophyte, of which 357 bp was exon sequence and the remainder was intron sequence. For phylogenetic analysis of the hosts, sequences for both cpDNA intergenic regions ({\em trnT}-{\em trnL} and {\em trnL}-{\em trnF}) and the {\em trnL} intron were aligned individually then concatenated to give a combined alignment of approximately 2200 bp. Analysis was also performed using the sequences from {\em tubB} and {\em tefA} separately.

\begin{figure}[!htp]
\begin{center}
\includegraphics[width=15cm]{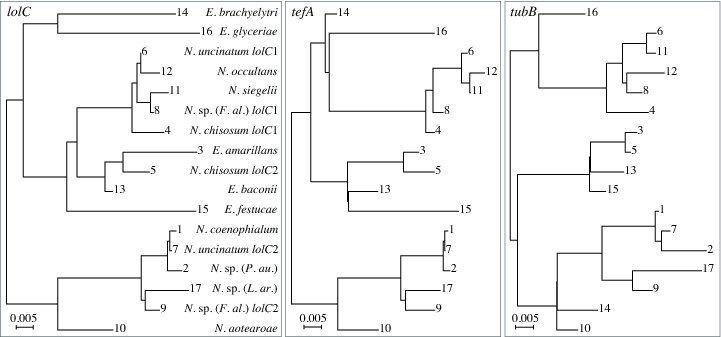}
\end{center}
\caption{\small Trees with maximum likelihood identified by MCMC search on aligned intron sequences from the {\em lolC, tefA,} and  {\em tubB} genes of {\em Epichloe} and {\em Neotyphodium} species. Some {\em Neotyphodium} species are interspecific hybrids that have multiple genomes from different ancestors. The genes from the different genomes are distinguished, for example, as {\em lolC}1 and  {\em lolC}2. The same number labeling leaves on the three trees indicates genes from the same genome from the same fungal isolate.}\label{endophytes} 
\end{figure}

The posterior distributions were estimated using MrBayes with the following parameters: (1) for the model: GTR + Gamma + Invariant sites; (2)
for MCMC:  number of runs: 1, number of chains: $2$, chain length: $100,000$, sample frequency: $1,000$, burn-in: $25$\%; and (3)
for bootstrap sampling: $100$ bootstrap samples, number of bootstrap columns equals length of original alignment.

%\begin{table}[!htp]
%\begin{center}
%\begin{tabular}{l|c}
%Data set& p-value  \\
%\midrule
%Grass-endophyte {\em tefA} & 0.040\\
%Grass-endophyte {\em tubB} & 0.080\\
%Grass-endophyte {\em tubB} plus {\em tefA} & 0.000\\
%\end{tabular}
%\end{center}
%\label{PE}  
%\caption{This table shows the p-values for the hypotheses with our statistical Method with the grass-endophyte data sets from \cite{chris}.  After removing cases of apparent host jumps, the data sets contain sequences from 20 taxa of grasses and 20 taxa of endophytes.   The first row represents for the endophyte data sets from {\em tefA} and the second row is from {\em tubB}.  The last row represents for the data sets with a concatenated sequences from {\em tubB} and {\em tefA}.}
%\end{table}

These results are interesting in comparison with the prior finding of significant relationship between the phylogenies of the grasses and their endophytes  \cite{chris}. The previous analysis indicated a significant relationship between ages of corresponding nodes in endophyte and grass phylogenies, addressing whether divergences of grass and endophyte clades tended to occur at approximately the same time. In contrast, results of the analysis above suggest that the grass and endophyte phylogenies are significantly different (Table \ref{PE}). We conclude that such a relationship of node ages does not necessarily imply similar phylogenetic histories. This is reasonable because the relationships of grasses and their endophytes is expected to be one of diffuse cospeciation at best. Individual species of endophyte may be associated with genera or tribes of grasses, but rarely with individual species. This contrasts with the gopher-gopher louse situation, where evidence suggests a much stricter coevolutionary relationship (Table \ref{GL}).

We chose an additional biological data set to compare phylogenies of genes that occur together in endophyte genomes. Whereas {\em tefA} and {\em tubB} are housekeeping genes present in all isolates, {\em lolC} is a secondary metabolism gene sporadically present in endophyte isolates \cite{Spiering2002}. It has been suggested that such sporadically occurring secondary metabolism genes may be distributed in fungi largely by horizontal gene transfer \cite{Walton2000}. To investigate this possibility in the case of {\em lolC}, we used our approach to test whether the phylogenies of these three genes were significantly different. The most likely trees obtained by MCMC showed related but nonidentical topologies (Figure \ref{endophytes}; note placement of genes from {\em Epichloe festucae} and {\em Epichloe brachyelytri}). Our test found no significant difference between the phylogenies, although the p-values appear stochastically smaller than the p-values observed for simulated data under the null.  This perhaps reflects the conservative nature of our test.  Removing either {\em Epichloe festucae} or {\em Epichloe brachyelytri}) altered the results only slightly (Table \ref{16taxa}). These results indicate that {\em lolC} evolution was largely or exclusively by decent, and disfavored horizontal transfer as an explanation for the sporadic distribution of this gene.

\begin{table}[!htp]
\begin{center}
\subtable%[TABBOTCAP]
[The results with our statistical method with the endophyte data sets from {\em lolC}, {\em tubB}, {\em tefA} genes.  There are 17 taxa in each data set.]{
$\phantom{AAAA}$
\begin{tabular}{l|c}
Data set& p-value  \\
\midrule
{\em lolC}  vs. {\em tefA} & 0.390\\
{\em lolC} vs. {\em tubB} & 0.560\\
{\em tefA} vs. {\em tubB} & 0.940\\
\end{tabular}
$\phantom{AAAA}$
%\end{center}
\label{tefAtubBlolC}  
%\caption{The results with our statistical method with the endophyte data sets from lolC, tubB, tefA genes.  There are 17 taxa in each data set.}
}
$\phantom{A}$
%\end{table}
%
%\begin{table}[!htp]
%\begin{center}
\subtable[The results with our statistical method with the endophyte data sets from {\em lolC}, {\em tubB}, {\em tefA} genes after removing E2368. There are 16 taxa in each data set.]{
%The results with our statistical method with the endophyte data sets from {\em lolC}, {\em tubB}, {\em tefA} genes.  There are 16 taxa in each data set after removing E2368 (there are 16 taxa in each data set). ]{
$\phantom{AAAA}$
\begin{tabular}{l|c}
Data set& p-value  \\
\midrule
{\em lolC}  vs. {\em tefA} & 0.230\\
{\em lolC} vs. {\em tubB} & 0.340\\
{\em tefA} vs. {\em tubB} & 0.870\\
\end{tabular}
$\phantom{AAAA}$
\label{16taxa}  
%\caption{The results with our statistical method with the endophyte data sets from lolC, tubB, tefA genes.  There are 17 taxa in each data set after removing E2368 (there are 16 taxa in each data set). }
}
\caption{}
\end{center}
\end{table}

\subsection{Toolkit for Computational Experiments}
To facilitate computations for our experiments, we developed a set of programs, collectively called {\tt Phylotree}. {\tt Phylotree} is organized as a collection of scripts for running a complete computational experiment starting from sequence alignments, then sampling phylogenetic trees and computing distances between phylogenetic trees and their distributions (see Section \ref{Methods}). Supported distance measures include path-difference, dissimilarity map distance, Robinson-Foulds distance. Available scripts allow for selecting the number of columns and the number of bootstrap samples, linking taxa in the alignments and provide flexibility for using different sampling methods (e.g., {\tt MrBayes} or {\tt BEAST}) and distance measures. This is free software, and will be distributed under the terms of the GNU General Public License.  One can download the software at \url{http://csurs7.csr.uky.edu/phylotree/}.  The website is password protected and login information can be obtained at \url{http://cophylogeny.net/research.php}.

\section{Discussion}
In this paper we presented a method to determine if two phylogenetic trees with given alignments are significantly incongruent.  Our method computes the difference of means of posterior distributions of trees, which has the advantage of using entire tree distributions, as opposed to single tree estimators.  

% We demonstrated our method with simulated data sets generated by {\tt Mesquite}.    

In this paper we used the triangle inequality ($d_1 < d_2 + d_3$ in Figure 1) to derive a bootstrap procedure to compute p-values.  However, our bootstrap procedure appears to be very conservative, producing p-values whose null distribution is stochastically much larger than uniform $U(0,1)$.  Thus in order to increase the power we might want to consider different criteria for computing p-values.  One approach may be to define $v(\hat{T_1}), v(\hat{T_2})$ to be the average of bootstraps $\{v(T_1^*)\}, \{v(T_2^*)\}$, rather than the initial tree estimates.  Another possibility is to replace the triangle inequality with a max condition (e.g. in Figure 1 use the condition $d_1 < \max (d_2, d_3)$).  We explored this in the supplementary material, and it seems that the max condition provides much more power, but is somewhat anti-conservative.
 
%One way to increase power is to compute the mean of the bootstrap means.  Namely, we compute $d_1$ in Figure 1.  Then we compute the mean of the means for the posterior distributions given hypothetical-$D$.  Then we compute the distance from the mean of the means for posterior distribution and a mean of a posterior distribution given a hypothetical-$D$, say $d_2$   Then we do this for $D'$, say $d_3$.   Then we estimate the p-value by using the cut-off $d_1 < d_2 + d_3$. 

In this paper we used the dissimilarity map as feature space.  However, there other common tree features which can be used to define different feature spaces. Examples of distances derived from tree features include (normalized) Robinson-Foulds distance \citep{Robinson1981}; quartet distance \citep{Estabrook1985}; and the path difference metric \citep{Steel1993}.
Of course, in all the above examples, we could choose any vector space norm, such as $L_p$ for any $p$.  The important point is that there are many different useful {\em features} (i.e., choices of vector space embeddings) which can be used to analyze trees, and many such as splits and quartets have already been used for quite some time. Moreover, with the kernel trick presented above we can efficiently calculate distances between distributions of trees using the Robinson-Foulds and quartet distance. Thus it is interesting to use different feature spaces for our statistical method, and we leave this for future work.

\section{Acknowledgments}
D. Haws, E. Arnaoudova, J. Jaromczyk, C. Schardl, R. Yoshida are supported by NIH R01 grant 5R01GM086888. E. Arnaoudova, J. Jaromczyk, and N. Moore developed the software {\tt Phylotree}.
P. Huggins is supported by the Lane Fellowship in Computational Biology at Carnegie Mellon University. 

\section{Supplement Materials}

Supplement materials can be downloaded at \url{http://cophylogeny.net/research.php}.

\bibliographystyle{elsart-harv}
\bibliography{ismb}

\pagebreak
\clearpage
\section{Supplement Materials}

\subsection{Bootstrap procedure}

Here is the outline of our bootstrap procedure we have used in the main manuscript.

\begin{algorithm} \label{algorithm}

{\bf Input}: Alignments $D_1$ and $D_2$

{\bf Output}: An estimated p-value, under the null $H_0:  ||v(T_1) - v(T_2)|| = 0$

\begin{enumerate}
\item Compute tree estimates $v(\hat{T_1}), v(\hat{T_2})$ from the data $D_1, \,D_2$. 
\item Compute $d = || v(\hat{T_1}) - v(\hat{T_2})||$. 
\item Let $p = 0$ and for $i = 1$ to $N$ do
\begin{itemize}
\item Take a bootstrap sample $D_1^*$ from the columns of alignment $D_1$ and a bootstrap sample ${D_2}^*$ from the columns of $D'$.
\item Compute tree estimates $v({T_1}^*), v({T_2}^*)$ from the data $D_1^*, {D_2}^*$.
\item Test the condition:  if $d \leq || v(\hat{T_1}) - v({T_1}^*)|| + || v(\hat{T_2}) - v({T_2}^*)||$ then set $p = p+1$. 
\end{itemize}
\item Set $p = \frac{p}{N}$ and return $p$ as the p-value of the hypothesis test.
\end{enumerate}
\end{algorithm}

During the ``Test the condition'' step, one could alternatively use the less conservative condition $d \leq \max (|| v(\hat{T_1}) - v({T_1}^*)|| , || v(\hat{T_2}) - v({T_2}^*)||)$

\subsection{Proof of  Proposition \ref{prop1}}

\begin{proof}   
For any $a \in \R^m$, let $a^T$ be the transpose of $a$.
Since $x_1, x_2, y_1, y_2 \in {\mathbb R}^m$ are mutually independent,
the bilinearity of dot-product gives ${\mathbb E}(x_1^T x_2) = {\mathbb E}(x_1)^T  {\mathbb E}(x_2)$ 
and similarly ${\mathbb E}(y_1^T y_2) = {\mathbb E}(y_1)^T {\mathbb E}(y_2)$ and ${\mathbb E}(x_1^T y_1) = {\mathbb E}(x_1)^T {\mathbb E}(y_1)$.
Also for any vectors $a, b \in \R^m$, we have the identity $a^T b = b^T a$.

Then we have
\begin{equation}\label{diffmeans11}
\begin{array}{ll}
&{\mathbb E}( || x_1 - y_1 ||^2 )\\
= & {\mathbb E}( (x_1 - y_1)^T (x_1 - y_1) )\\
= & {\mathbb E}(x_1^T x_1 +  y_1^T y_1 - x_1^T y_1 - y_1^T x_1)\\
= & {\mathbb E}(x_1^T x_1) + {\mathbb E}(y_1^T y_1) - {\mathbb E}(x_1^T y_1) - {\mathbb E}(y_1^T x_1)\\
= & {\mathbb E}(x_1^T x_1) + {\mathbb E}(y_1^T y_1) - {\mathbb E}(x_1)^T{\mathbb E}(y_1) - {\mathbb E}(y_1)^T {\mathbb E}(x_1)\\
= & {\mathbb E}( x_1^T x_1) + {\mathbb E}( y_1^T y_1) - 2\mu_x^T \mu_y.\\
\end{array}
\end{equation}

But
\begin{equation*}\label{diffmeans22}
\begin{array}{ll}
&{\mathbb E}(x_1^Tx_1) \\
= & {\mathbb E}\left[(x_1 - x_2 + x_2)^T(x_1 - x_2 + x_2)\right]\\
= & {\mathbb E}[(x_1 - x_2)^T(x_1 - x_2)+x_2^T(x_1 - x_2)\\
&+(x_1 - x_2)^Tx_2 + x_2^Tx_2]\\
= & {\mathbb E}\left[(x_1 - x_2)^T(x_1 - x_2)\right]+{\mathbb E}\left[x_2^T(x_1 - x_2)\right]\\
&+{\mathbb E}\left[(x_1 - x_2)^Tx_2\right] + {\mathbb E}(x_2^Tx_2)\\
= & {\mathbb E}\left[(x_1 - x_2)^T(x_1 - x_2)\right]+{\mathbb E}(x_2^Tx_1) -{\mathbb E}(x_2^T x_2)\\
&+{\mathbb E}(x_1^T x_2) - {\mathbb E}(x_2^Tx_2) + {\mathbb E}(x_2^Tx_2)\\
= & {\mathbb E}\left[(x_1 - x_2)^T(x_1 - x_2)\right]+{\mathbb E}(x_2^Tx_1) -{\mathbb E}(x_2^T x_2)\\
&+{\mathbb E}(x_1^T x_2)\\
= & {\mathbb E}\left[(x_1 - x_2)^T(x_1 - x_2)\right] + \mu_x^T \mu_x - {\mathbb E}(x_2^Tx_2) \\
&+ \mu_x^T \mu_x\\
= & {\mathbb E}\left[(x_1 - x_2)^T(x_1 - x_2)\right] + 2\mu_x^T \mu_x - {\mathbb E}(x_2^Tx_2)\\
= & {\mathbb E}\left[(x_1 - x_2)^T(x_1 - x_2)\right] + 2\mu_x^T \mu_x - {\mathbb E}(x_1^Tx_1)\\
\end{array}
\end{equation*}
Thus we have
\begin{equation}\label{diffmeans33}
 2 {\mathbb E}(x_1^Tx_1)
= {\mathbb E}\left[(x_1 - x_2)^T(x_1 - x_2)\right] + 2\mu_x^T \mu_x .
\end{equation}
By dividing the both sides of the equation in \eqref{diffmeans33} by 2 we have:
\begin{equation}\label{diffmeans44}
 {\mathbb E}(x_1^Tx_1)
=\frac{1}{2} {\mathbb E}\left[(x_1 - x_2)^T(x_1 - x_2)\right] + \mu_x^T \mu_x .
\end{equation}
Similarly we have
\begin{equation}\label{diffmeans55}
 {\mathbb E}(y_1^Ty_1)
=\frac{1}{2} {\mathbb E}\left[(y_1 - y_2)^T(y_1 - y_2)\right] + \mu_y^T \mu_y .
\end{equation}
Then we substitute ${\mathbb E}(x_1^Tx_1)$ and ${\mathbb E}(y_1^Ty_1)$ in equations \eqref{diffmeans44} and \eqref{diffmeans55} into the
equation in \eqref{diffmeans11}, we have
 \begin{equation}\label{diffmeans66}
\begin{array}{ll}
&{\mathbb E}( || x_1 - y_1 ||^2 )\\
= &\left[ \frac{1}{2} {\mathbb E}\left[(x_1 - x_2)^T(x_1 - x_2)\right] + \mu_x^T \mu_x\right]\\
& + \left[\frac{1}{2} {\mathbb E}\left[(y_1 - y_2)^T(y_1 - y_2)\right] + \mu_y^T \mu_y\right]\\
& - 2\mu_x^T \mu_y\\
=& \frac{1}{2} {\mathbb E}\left[(x_1 - x_2)^T(x_1 - x_2)\right] + \frac{1}{2} {\mathbb E}\left[(y_1 - y_2)^T(y_1 - y_2)\right]\\
& + \left[\mu_x^T \mu_x+\mu_y^T\mu - 2\mu_x^T \mu_y \right]\\
=& \frac{1}{2} {\mathbb E}\left[(x_1 - x_2)^T(x_1 - x_2)\right] + \frac{1}{2} {\mathbb E}\left[(y_1 - y_2)^T(y_1 - y_2)\right]\\
& + ||\mu_x + \mu_y||^2.\\
\end{array}
\end{equation}
\end{proof}

\subsection{Supplement for simulation studies}

Throughout the remaining subsections, Condition (i) refers to the condition tested in Algorithm \ref{algorithm}: $d \leq || v(\hat{T_1}) - v({T_1}^*)|| + || v(\hat{T_2}) - v({T_2}^*)||$.  Condition (ii) refers to the alternative condition  $d \leq \max (|| v(\hat{T_1}) - v({T_1}^*)|| , || v(\hat{T_2}) - v({T_2}^*)||)$.

\begin{sidewaystable}[!htp]
\begin{center}
\begin{tabular}{l|ccc|ccc|ccc}
 & \multicolumn{3}{c|}{1000K SD} & \multicolumn{3}{|c|}{600K SD} & \multicolumn{3}{|c}{100K SD}\\
Data set & i & ii & dist & i & ii & dist & i & ii & dist  \\
\midrule
Sp1\_Gene0\_rep1 / Sp1\_Gene0\_rep2 & 1.000 &  0.790 &  0.046  & 0.980 &  0.140 &  0.071 &  0.870 &  0.100 &  0.142\\
Sp1\_Gene1\_rep1 / Sp1\_Gene1\_rep2 & 0.990 &  0.430 &  0.066  & 0.990 &  0.800 &  0.051 &  0.870 &  0.180 &  0.116\\
Sp1\_Gene2\_rep1 / Sp1\_Gene2\_rep2 & 0.330 &  0.000 &  0.105  & 1.000 &  0.440 &  0.061 &  1.000 &  0.700 &  0.081\\
Sp1\_Gene3\_rep1 / Sp1\_Gene3\_rep2 & 0.980 &  0.340 &  0.063  & 0.760 &  0.000 &  0.092 &  0.700 &  0.040 &  0.136\\
Sp1\_Gene4\_rep1 / Sp1\_Gene4\_rep2 & 0.940 &  0.110 &  0.072  & 0.990 &  0.370 &  0.063 &  1.000 &  0.410 &  0.103\\
Sp1\_Gene5\_rep1 / Sp1\_Gene5\_rep2 & 0.550 &  0.050 &  0.090  & 0.750 &  0.040 &  0.087 &  0.810 &  0.030 &  0.134\\
Sp1\_Gene6\_rep1 / Sp1\_Gene6\_rep2 & 0.710 &  0.020 &  0.091  & 0.990 &  0.730 &  0.055 &  1.000 &  0.800 &  0.090\\
Sp1\_Gene7\_rep1 / Sp1\_Gene7\_rep2 & 0.940 &  0.150 &  0.077  & 1.000 &  0.600 &  0.057 &  0.570 &  0.010 &  0.159\\
Sp1\_Gene8\_rep1 / Sp1\_Gene8\_rep2 & 0.980 &  0.340 &  0.061  & 0.970 &  0.110 &  0.067 &  0.950 &  0.570 &  0.083\\
Sp1\_Gene9\_rep1 / Sp1\_Gene9\_rep2 & 1.000 &  0.580 &  0.058  & 0.780 &  0.040 &  0.090 &  0.920 &  0.160 &  0.112\\
\midrule
\end{tabular}
\end{center}
\caption{Testing Type I error by comparing the identical gene trees generated by {\tt Mesquite}.  We tested ten gene trees for each species depth (i.e. 30 different gene trees in total) generated under the same species tree.  We used two sets of sequences generated under the HKY model with the same tree for each test.  The numbers in the table are p-value for our hypothesis testing.  We give p-values for the three species depths of 1000,000, 600,000 and 100,000, with fixed population size of 100,000.  We used the species tree on right with each species depth in Figure \ref{species}.  dist represents the average of the normalized difference of means with posterior distributions given the original sequence data sets.}\label{typeI}  
\end{sidewaystable}

\begin{sidewaystable}[!htp]
\begin{center}
\begin{tabular}{l|cccc|cccc|cccc}
 & \multicolumn{4}{c|}{1000K SD} & \multicolumn{4}{|c|}{600K SD} & \multicolumn{4}{|c}{100K SD}\\
Data set & \Side{ i } & \Side{ ii } & \Side{ dist} & \Side{true dist} & \Side{ i } & \Side{ ii } & \Side{ dist} & \Side{true dist} & \Side{ i, } & \Side{ ii, } & \Side{ dist} & \Side{ true dist}   \\
\midrule
Sp1\_Genetree0 / Sp1\_Genetree1  & $ 0.910 $  &  $  0.070 $  &  $  0.079 $  &  $  0.044 $  &  $  1.000 $  &  $  0.880 $  &  $  0.047 $  &  $  0.044 $  &  $  0.000 $  &  $  0.000 $  &  $  0.284 $  &  $  0.290 $ \\
Sp1\_Genetree0 / Sp1\_Genetree2  & $ 0.640 $  &  $  0.010 $  &  $  0.093 $  &  $  0.060 $  &  $  0.970 $  &  $  0.320 $  &  $  0.067 $  &  $  0.039 $  &  $  0.120 $  &  $  0.000 $  &  $  0.205 $  &  $  0.182 $ \\
Sp1\_Genetree0 / Sp1\_Genetree3  & $ 0.230 $  &  $  0.000 $  &  $  0.114 $  &  $  0.074 $  &  $  0.840 $  &  $  0.040 $  &  $  0.088 $  &  $  0.056 $  &  $  0.950 $  &  $  0.210 $  &  $  0.118 $  &  $  0.172 $ \\
Sp1\_Genetree0 / Sp1\_Genetree4  & $ 0.310 $  &  $  0.010 $  &  $  0.105 $  &  $  0.083 $  &  $  1.000 $  &  $  0.710 $  &  $  0.055 $  &  $  0.051 $  &  $  0.290 $  &  $  0.000 $  &  $  0.204 $  &  $  0.164 $ \\
Sp1\_Genetree0 / Sp1\_Genetree5  & $ 0.320 $  &  $  0.000 $  &  $  0.111 $  &  $  0.082 $  &  $  0.380 $  &  $  0.000 $  &  $  0.107 $  &  $  0.088 $  &  $  0.310 $  &  $  0.000 $  &  $  0.192 $  &  $  0.229 $ \\
Sp1\_Genetree0 / Sp1\_Genetree6  & $ 0.370 $  &  $  0.000 $  &  $  0.101 $  &  $  0.055 $  &  $  0.940 $  &  $  0.120 $  &  $  0.078 $  &  $  0.039 $  &  $  0.980 $  &  $  0.490 $  &  $  0.107 $  &  $  0.096 $ \\
Sp1\_Genetree0 / Sp1\_Genetree7  & $ 0.000 $  &  $  0.000 $  &  $  0.208 $  &  $  0.159 $  &  $  0.980 $  &  $  0.200 $  &  $  0.076 $  &  $  0.049 $  &  $  0.190 $  &  $  0.000 $  &  $  0.207 $  &  $  0.136 $ \\
Sp1\_Genetree0 / Sp1\_Genetree8  & $ 0.640 $  &  $  0.020 $  &  $  0.087 $  &  $  0.078 $  &  $  0.170 $  &  $  0.000 $  &  $  0.116 $  &  $  0.128 $  &  $  0.050 $  &  $  0.000 $  &  $  0.240 $  &  $  0.269 $ \\
Sp1\_Genetree0 / Sp1\_Genetree9  & $ 0.640 $  &  $  0.020 $  &  $  0.093 $  &  $  0.079 $  &  $  0.620 $  &  $  0.000 $  &  $  0.101 $  &  $  0.054 $  &  $  0.070 $  &  $  0.000 $  &  $  0.236 $  &  $  0.270 $ \\
Sp1\_Genetree0 / Sp2\_Genetree1  & $ 0.000 $  &  $  0.000 $  &  $  0.192 $  &  $  0.202 $  &  $  0.000 $  &  $  0.000 $  &  $  0.194 $  &  $  0.176 $  &  $  0.160 $  &  $  0.000 $  &  $  0.218 $  &  $  0.248 $ \\
Sp1\_Genetree0 / Sp2\_Genetree2  & $ 0.000 $  &  $  0.000 $  &  $  0.207 $  &  $  0.245 $  &  $  0.000 $  &  $  0.000 $  &  $  0.253 $  &  $  0.252 $  &  $  0.480 $  &  $  0.000 $  &  $  0.181 $  &  $  0.156 $ \\
Sp1\_Genetree0 / Sp2\_Genetree3  & $ 0.000 $  &  $  0.000 $  &  $  0.180 $  &  $  0.209 $  &  $  0.000 $  &  $  0.000 $  &  $  0.193 $  &  $  0.188 $  &  $  0.120 $  &  $  0.000 $  &  $  0.234 $  &  $  0.256 $ \\
Sp1\_Genetree0 / Sp2\_Genetree4  & $ 0.000 $  &  $  0.000 $  &  $  0.237 $  &  $  0.269 $  &  $  0.000 $  &  $  0.000 $  &  $  0.199 $  &  $  0.232 $  &  $  0.050 $  &  $  0.000 $  &  $  0.244 $  &  $  0.237 $ \\
Sp1\_Genetree0 / Sp2\_Genetree5  & $ 0.000 $  &  $  0.000 $  &  $  0.213 $  &  $  0.233 $  &  $  0.000 $  &  $  0.000 $  &  $  0.281 $  &  $  0.283 $  &  $  0.290 $  &  $  0.000 $  &  $  0.202 $  &  $  0.205 $ \\
Sp1\_Genetree0 / Sp2\_Genetree6  & $ 0.000 $  &  $  0.000 $  &  $  0.192 $  &  $  0.245 $  &  $  0.000 $  &  $  0.000 $  &  $  0.259 $  &  $  0.236 $  &  $  0.080 $  &  $  0.000 $  &  $  0.234 $  &  $  0.276 $ \\
Sp1\_Genetree0 / Sp2\_Genetree7  & $ 0.000 $  &  $  0.000 $  &  $  0.218 $  &  $  0.253 $  &  $  0.000 $  &  $  0.000 $  &  $  0.225 $  &  $  0.231 $  &  $  0.000 $  &  $  0.000 $  &  $  0.308 $  &  $  0.283 $ \\
Sp1\_Genetree0 / Sp2\_Genetree8  & $ 0.000 $  &  $  0.000 $  &  $  0.177 $  &  $  0.213 $  &  $  0.000 $  &  $  0.000 $  &  $  0.225 $  &  $  0.232 $  &  $  0.150 $  &  $  0.000 $  &  $  0.222 $  &  $  0.218 $ \\
Sp1\_Genetree0 / Sp2\_Genetree9  & $ 0.000 $  &  $  0.000 $  &  $  0.209 $  &  $  0.235 $  &  $  0.000 $  &  $  0.000 $  &  $  0.225 $  &  $  0.213 $  &  $  0.460 $  &  $  0.000 $  &  $  0.188 $  &  $  0.259 $ \\
\bottomrule
\end{tabular}
\end{center}
\caption{Shown are the p-values calculated using our method on eighteen pairs of gene trees.  The first nine lines  compare different genes derived from the same species tree (the species tree on left in each species depth in Figure \ref{species}), the second nine lines compare two genes from different species trees. We give p-values for the three species depths of 1,000,000, 600,000 and 100,000, with fixed population size of 100,000.  dist represents the average of the normalized difference of means with posterior distributions given the original sequence data sets and true dist represents the normalized distance between true trees.}\label{typeII}  
\end{sidewaystable}

\subsection{Results with biological data sets}

\begin{table}[!htp]
\begin{center}
\begin{tabular}{l|ccc}
Data set & \Side{ i } & \Side{ii} & \Side{ dist} \\
\midrule
Gopher-louse (dataset 1)        &  $ 0.640 $   &  $ 0.020 $  & $   0.120 $ \\
Gopher-louse (dataset 2)        &  $ 0.400 $   &  $ 0.020 $  & $   0.128 $ \\
Gopher-louse (dataset 3)        &  $ 0.840 $   &  $ 0.070 $  & $   0.117 $ \\
Gopher-louse (dataset 4)        &  $ 0.590 $   &  $ 0.030 $  & $   0.122 $ \\
\midrule
Grass-endophyte ({\em tefA} gene)        & $  0.040 $  & $ 0.000 $ &  $ 0.073$ \\
Grass-endophyte ({\em tubB} gene)        & $  0.080 $  & $ 0.010 $ &  $ 0.088$ \\
Grass-endophyte ({\em tefA}\_{\em tubB} genes) & $  0.000 $  & $ 0.000 $ &  $ 0.073$ \\
\bottomrule
\end{tabular}
\end{center}
\caption{The numbers in this table are p-values for our statistical hypothesis testing estimated by our method with the grass-endophyte data sets from \cite{chris}.  dist represents the average of the normalized difference of means with posterior distributions given the original sequence data sets.}\label{PE2}  
\end{table}

\begin{table}[!htp]
\begin{center}
\begin{tabular}{l|cccc}
Data set & \Side{ i } & \Side{ii} & \Side{Permuted} & \Side{ dist} \\
\midrule
(- E2368)	\\
endophyte 16-taxa (lolC vs tubB)&  0.340&  0.000&  0.000 & 0.100\\
endophyte 16-taxa (lolC vs tefA)&  0.230&  0.000&  0.000 & 0.105\\
endophyte 16-taxa (tefA vs tubB)&  0.870&  0.050&  0.000 & 0.080\\
\midrule
(- E1125) \\
endophyte 16-taxa (lolC vs tubB)& 0.670& 0.010& &  0.093\\
endophyte 16-taxa (lolC vs tefA)& 0.870& 0.040& &  0.073\\
endophyte 16-taxa (tefA vs tubB)& 0.950& 0.160& &  0.069\\
\bottomrule
\end{tabular}
\end{center}
\caption{The numbers in this table are p-values for our statistical hypothesis testing estimated by our method with the endophyte data sets from {\em tefA}, {\em tubB}, and {\em lolC} genes.  dist represents the average of the normalized difference of means with posterior distributions given the original sequence data sets.  (- E2368) means that they are with the data sets after removing E2368 and (- E1125) means that these are the results with the data sets after removing E1125.}\label{lolC}  
\end{table}

\end{document}